\documentclass{article}
\usepackage{spconf,graphicx,hyperref}
\usepackage{cite}
\usepackage{amsmath,amssymb,amsfonts}
\usepackage{algorithmic}
\usepackage{textcomp}
\usepackage{xcolor}
\usepackage{multirow}
\usepackage{booktabs}
\usepackage{cleveref}

\hypersetup{
  colorlinks=true,    
  linkcolor=black,     
  citecolor=black,    
  urlcolor=black 
}

\def\BibTeX{{\rm B\kern-.05em{\sc i\kern-.025em b}\kern-.08em
    T\kern-.1667em\lower.7ex\hbox{E}\kern-.125emX}}
\newcommand{\spm}[1]{{\footnotesize \textpm #1}}

\title{%
    STAR: Speech-to-Audio Generation via Representation Learning\\}
    
\name{Zeyu Xie$^{1}$, Xuenan Xu$^{2}$, Yixuan Li$^{2}$, Mengyue Wu$^{2}$\textsuperscript{*}, Yuexian Zou$^{1}$\textsuperscript{*}\thanks{* Corresponding authors}}

\address{$^{1}$ Guangdong Provincial Key Laboratory of Ultra High Definition Immersive Media Technology, \\
Peking University, Shenzhen \\
         $^{2}$ X-LANCE Lab, Shanghai Jiao Tong University, Shanghai \\
         \textit{zeyuxie25@stu.pku.edu.cn}, \textit{zouyx@pku.edu.cn}}

\begin{document}
%
\maketitle
\begin{abstract}
This work presents STAR, the first end-to-end speech-to-audio generation framework, designed to enhance efficiency and address error propagation inherent in cascaded systems.
Unlike prior approaches relying on text or vision, STAR leverages speech as it constitutes a natural modality for interaction.
As an initial step to validate the feasibility of the system, we demonstrate through representation learning experiments that spoken sound event semantics can be effectively extracted from raw speech,  capturing both auditory events and scene cues.
Leveraging the semantic representations, STAR incorporates a bridge network for representation mapping and a two-stage training strategy to achieve end-to-end synthesis.
With a $76.9\%$ reduction in speech processing latency, STAR demonstrates superior generation performance over the cascaded systems.
Overall, STAR establishes speech as a direct interaction signal for audio generation, thereby bridging representation learning and multimodal synthesis.
Generated samples are available at \href{https://zeyuxie29.github.io/STAR/}{\textcolor{cyan}{\textit{https://zeyuxie29.github.io/STAR}}}.

\end{abstract}
\section{Introduction}
\label{sec:intro}

While significant progress in audio generation, speech — a natural and fundamental form of human communication and an interaction-oriented input modality — has received little attention. 
For instance, given a spoken query such as ``the sound of birds'', no current system can directly produce the corresponding audio.
To address this gap, we propose the speech-to-audio (STA) task, which aims to understand event information within speech and generate corresponding audio.

Naturally, one would argue that such task can be achieved by a cascaded system consisting of automatic speech recognition (ASR) and text-to-audio (TTA) models.
However, the cascaded system introduces unnecessary latency, which can significantly hinder the human-computer interaction experience.
Our analysis reveals that at least $156ms$ of delay is required to process a speech input in a cascaded system, while our end-to-end (E2E) approach reduces the latency to $36ms$ ($\approx 76.9\%$ reduction).
Furthermore, E2E models offer the key advantage of unified training with speech language models, enabling seamless integration of audio generation into human-computer interaction.
This unlocks new capabilities for systems like Speech SpeechGPT~\cite{zhang2023speechgpt} and Qwen-Omni~\cite{xu2025qwen2}, overcoming the fundamental limitations of cascaded systems.
By design, E2E systems deliver low latency, native integration, and reduced cascading errors, establishing a new track for audio language models.

\begin{figure}[t]
\centerline{\includegraphics[width=\linewidth]{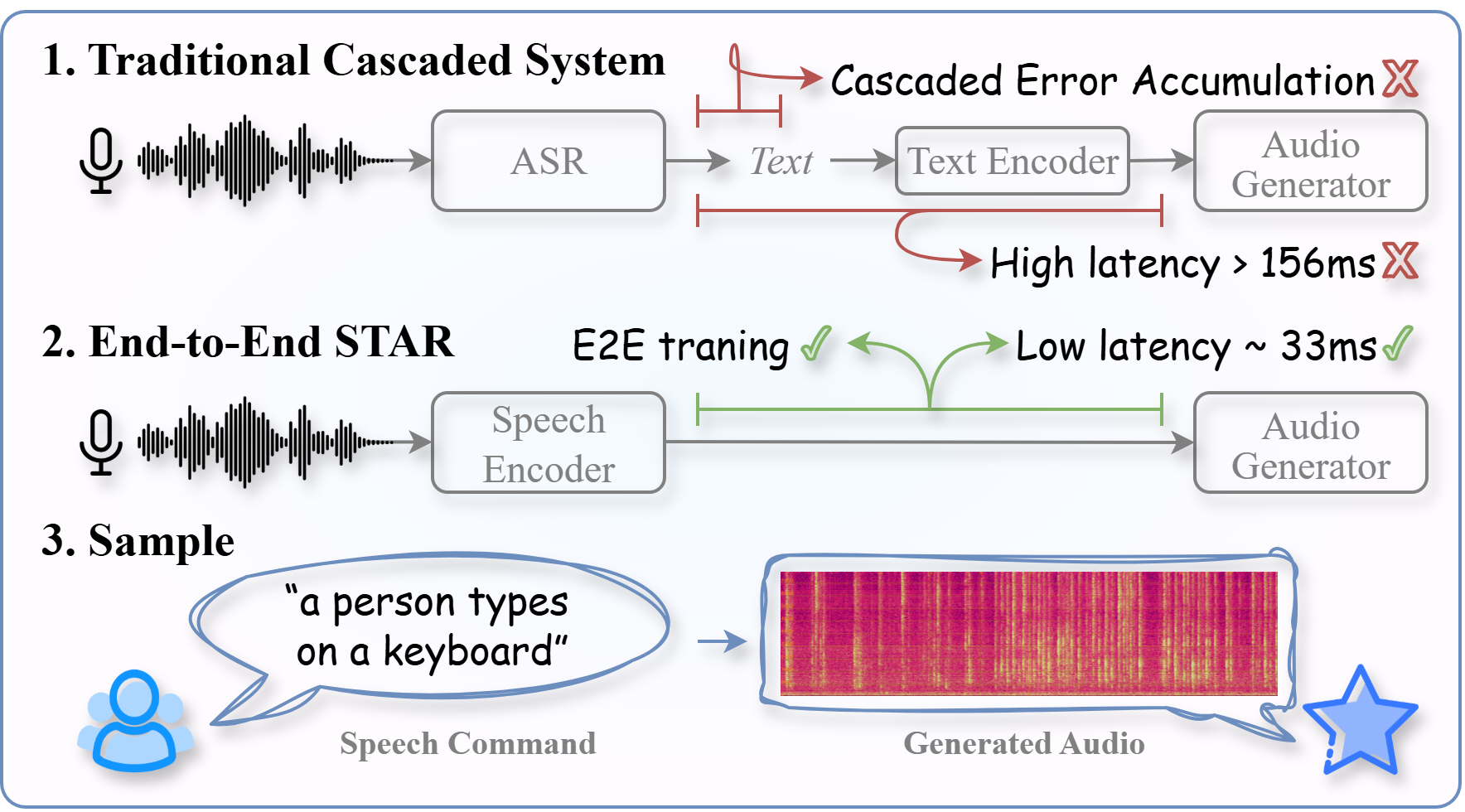}}
\caption{
The speech-to-audio model generates audio directly from user' speech, offering lower latency, reduced error propagation, and E2E integration compared to cascaded systems.
}
\label{fig}
\end{figure}

Another pertinent issue concerns the feasibility of the E2E STA system — namely, whether sufficient information can be \textbf{directly extracted} from speech signals to drive audio generation.
Specifically, can we extract speech representations that contain semantic information of sound events like ``dog barking'' or ``bell ringing''?
Such representations are referred to as \underline{spoken sound event semantics}.
To this end, we conducted a series of representation learning experiments to evaluate the cross-modal information extraction capability of different encoders.
The results suggest that self-supervised pre-training speech encoders, such as HuBERT~\cite{hsu2021hubert} and WavLM~\cite{chen2022wavlm}, \textbf{are capable of extracting spoken sound event semantics from speech signals}.
On the strength of this evidence, we developed STAR (\textcolor{gray!90}{S}peech-\textcolor{gray!90}{T}o-\textcolor{gray!90}{A}udio generation via \textcolor{gray!90}{R}epresentation learning).

To better bridge the speech and audio modalities, we draw on insights from large-model fine-tuning~\cite{chu2023qwen} to introduce a bridge network and a two-stage training strategy.
Specifically, our framework consists of a speech encoder for extracting speech embeddings, a bridge network that maps them to spoken sound event semantics, and a flow matching model for generation. 
The two-stage training comprises (1) representation learning for speech-audio semantic alignment, and (2) generative model training.
Experimental results demonstrate that our end-to-end system significantly reduces speech processing latency while surpassing the generation performance of cascaded systems.
This also validates the effectiveness of leveraging speech as a direct input modality to guide audio generation, making it a pioneering exploration for integration with audio LLMs.
Our contributions are as follows: 
\begin{enumerate}
    \item Recognize the potential of the speech-to-audio generation task and have designed the first E2E system STAR;
    \item Validate E2E STA feasibility via representation learning experiments, showing that spoken sound event semantics can be directly extracted;
    \item Achieve effective speech-to-audio modal alignment through a bridge network mapping mechanism and a two-stage training strategy;
    \item Significantly reduces speech processing latency from $156ms$ to $36ms$ ($\approx 76.9\%$ reduction), while surpassing the generation performance of cascaded systems.
    

\end{enumerate}

\begin{figure*}[t]
\centerline{\includegraphics[width=\textwidth]{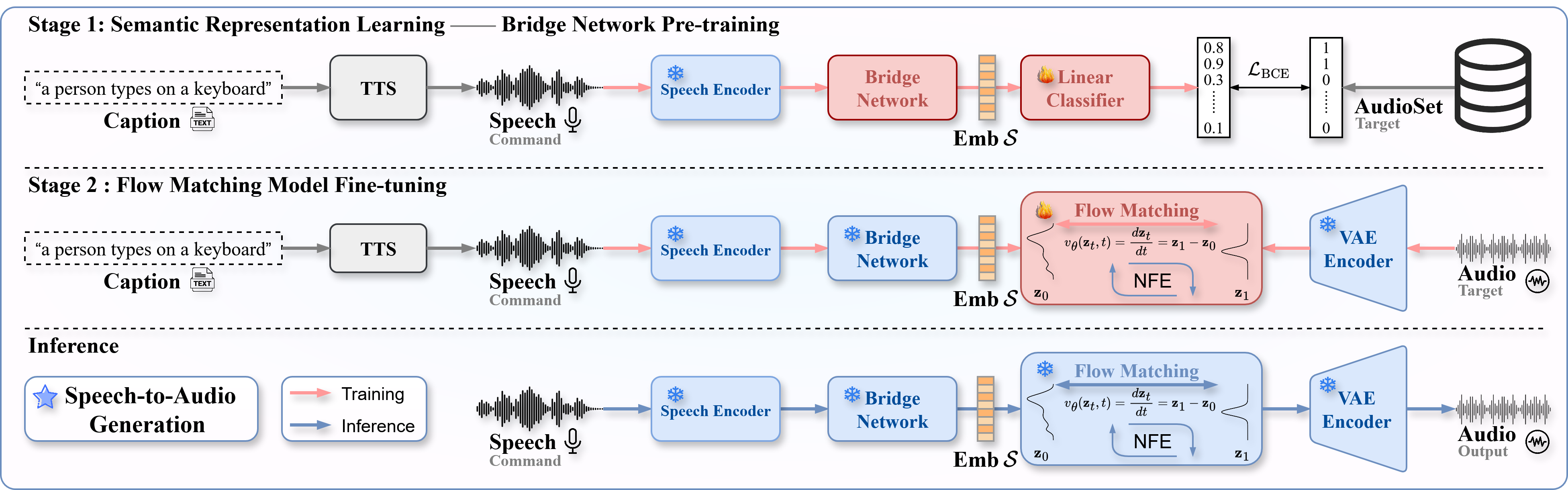}}
\caption{
An overview of STAR. 
For an input speech utterance, stage (1) leverages representation learning — validated through the sound event classification task — to demonstrate that spoken sound event semantics $\mathcal{S}$ can be effectively and directly derived, while simultaneously pre-training the bridge network.
Stage (2) fine-tunes the flow matching model based on stage (1); this model generates a latent representation conditioned on $\mathcal{S}$, which is then reconstructed into waveform by the VAE decoder.
}
\label{fig:pipeline}
\end{figure*}

\section{Spoken sound event semantics Representations Learning}
\label{sec:representations}

Representation learning was conducted to validate the feasibility of end-to-end STA, as it is essential for extracting sufficient semantic information from speech — particularly information pertaining to sound event types.
Hence we utilize a combination of speech encoder and bridge network to (1) extract high-level compact features from raw speech signals and (2) map features to spoken sound event semantic representation that
contain sound event information, respectively.

To facilitate the narrative, we first provide the following definitions:
A \underline{caption} refers to textual content that describes sound events. 
In STA, the input is the speech waveform that reads the caption out, hereafter referred to as \underline{speech}. 
The generated content is \underline{audio}.

\subsection{Speech Encoder}
A speech encoder is necessary to extract compact embedding from redundant raw speech signals.
We investigate the efficacy of different types of encoders: one type includes HuBERT~\cite{hsu2021hubert} and WavLM~\cite{chen2022wavlm}, which are pre-trained with masked language modeling objectives; the other type consists of codec models, such as the descript-audio-codec (DAC)~\cite{kumar2024high}, which is trained using an autoencoding approach.

\subsection{Bridge Network}
The extracted speech embedding is denoted as $\mathcal{\hat{S}}$.
To align $\mathcal{\hat{S}}$ with sound event semantic representations $\mathcal{S}$ for guiding audio generation, we employ a bridge network for mapping.
Two architectures are investigated:
(1) MLP directly projects $\mathcal{\hat{S}}$ by fully-connected layers and an average pooling operation;
(2) Q-Former~\cite{devlin2018bert} encodes $\mathcal{\hat{S}}$ through cross-attention with a fixed number of learnable query embeddings.

In our representation learning experiments, the downstream task of sound event classification is utilized to evaluate the quality of the mapped representation, since accurate sound event identification is paramount for audio generation.

\section{STAR: E2E Speech-to-Audio Generation}
The representation learning results (see Section~\ref{subsec:speech_rep_res}) demonstrate that spoken sound event semantics can be effectively and directly extracted from speech signals, thus enabling the development of an E2E STA system.
As shown in \Cref{fig:pipeline}, STAR combines the speech encoder and bridge network from representation experiments to extract semantic representations, and further incorporates a generation module composed of flow matching and a variational autoencoder (VAE).

\subsection{VAE for Audio Compression} The VAE is employed to extract representations from audio to reduce the computation burden.
The VAE encoder compresses the audio $\mathcal{A} \in \mathbb{R}^{T}$ into the latent representation $\mathbf{z} \in \mathbb{R}^{\frac{T}{R}\times D} = [\mathbf{z}_{\mu}, \mathbf{z}_{\sigma}]$ with a compression ratio $R$.
The VAE decoder reconstructs the audio $\mathcal{\tilde{A}}$ based on samples drawn from the distribution $\mathcal{N}(\mathbf{z}_{\mu}, \mathbf{z}_{\sigma})$.

\subsection{Flow Matching Model}
A flow matching model~\cite{liu2022flow} is utilized to predict $\mathbf{\tilde{z}}$ based on semantic feature $\mathcal{S}$. 
It defines a continuous-time flow from the data distribution $\mathbf{z}_0 \sim p_\text{data}$ to a simple prior distribution $\mathbf{z}_1 \sim \mathcal{N}(0, \mathbf{I})$. 
The flow is represented by a velocity field $v_\theta(\mathbf{z}_t, t)$ such that the intermediate latents follow
\begin{equation}
    \frac{d \mathbf{z}_t}{dt} = v_\theta(\mathbf{z}_t, t), \quad 
    \mathbf{z}_t = (1-t) \mathbf{z}_0 + t \mathbf{z}_1
\end{equation}
where $t \in [0,1]$, $\mathbf{z}_0$ is sampled from the data distribution and $\mathbf{z}_1$ from a standard Gaussian.  
The model is trained to minimize the flow matching loss, which enforces the velocity field to match the target:
\begin{equation}
  \label{eqn:flow_matching_loss}
  \mathcal{L} = \mathbb{E}_{t, \mathbf{z}_0, \mathbf{z}_1} \big\| v_\theta(\mathbf{z}_t, t, \mathcal{S}) - (\mathbf{z}_1 - \mathbf{z}_0) \big\|^2
\end{equation}
where $\theta$ denotes the model parameters, and the expectation is taken over uniformly sampled $t \in [0,1]$ and pairs $(\mathbf{z}_0, \mathbf{z}_1)$.  

During inference, an ordinary differential equation (ODE) solver is used to sample $\mathbf{\tilde{z}}_0$ given $\frac{d \mathbf{z}_t}{dt}$ estimated by the network.
We use Euler ODE solver with the sway sampling schedule for timesteps~\cite{chen2024f5} which has been demonstrated to improve performance when the number of function evaluations (NFE) is limited.

For the architecture, we adopt the Diffusion Transformer (DiT) with the AdaLN-SOLA step fusion mechanism as our backbone, as proposed by EzAudio~\cite{hai2024ezaudio}.
We use the DiT-L variant, comprising 24 DiT blocks with a hidden size of 1024.

\section{Experimental Setup}
\noindent\textbf{Datasets}
For an E2E STA system, we create the speech form of caption to construct training speech-audio pairs.
We use AudioCaps~\cite{kim2019audiocaps} and a text-to-speech model VITS~\cite{kim2021conditional} to convert caption into corresponding speech.
Since AudioCaps is a subset of the sound event classification dataset, AudioSet~\cite{gemmeke2017audio}, the event labels are accessible for representation learning.

\textbf{Stage 1: Semantic representation learning}
We conduct representation learning—while simultaneously pre-training the bridge network—by appending a classifier for sound event classification, using labels from AudioSet~\cite{gemmeke2017audio}.
The bridge network is pre-trained with a learning rate of $3.2 \times 10^{-3}$, a maximum of $500$ epochs and an early stopping of $20$ epochs.
The evaluation metric is mean average precision (mAP).

\textbf{Stage 2: Fine-tuning}
The DiT backbone is then fine-tuned on STA task for 100 epochs.
The backbone is initialized with the one pre-trained on TTA.
The AdamW optimizer is adopted with a learning rate of $1 \times 5^{-5}$. 
The bridge network and VAE~\cite{hai2024ezaudio} are frozen.



\begin{table}[b]
\renewcommand{\arraystretch}{1.1}
\footnotesize
\caption{Linear probing classification results on AudioCaps. 
The caption representation performance serves as the topline.
}
\begin{center}
\begin{tabular}{c|cc|ccc}
\toprule
\textbf{Input}&\multicolumn{2}{c|}{\textbf{Caption}} &\multicolumn{3}{c}{\textbf{Speech}}\\
\midrule
\textbf{Encoder} & CLAP & Flan-T5 & HuBERT & WavLM & DAC\\
\midrule
MLP & 0.500& 0.419 & \textbf{0.457} & 0.451  & 0.114\\
Q-Former & 0.472 & 0.486 & \textbf{0.457} & 0.453& 0.108\\
\bottomrule
\end{tabular}
\label{tab:hear}
\end{center}
\end{table}

\begin{table*}[htbp]
\small
\renewcommand{\arraystretch}{1}
\caption{Generation performance of different systems.
``Processing latency" refers to the representation processing delay for one speech input, with the results representing the mean and variance. 
OVL = overall quality. REL = relevance to input.}

\begin{center}
\begin{tabular}{l|c|c|ccccc|cc}
\toprule
{\multirow{2}{*}{\textbf{System}}} &\textbf{Processing } & \textbf{Generation}
  &\multicolumn{5}{c|}{\textbf{Objective}} &\multicolumn{2}{c}{\textbf{Subjective}}\\
  

&\textbf{Latency (ms)} $\downarrow$ &\textbf{Time (s)} $\downarrow$ &FAD $\downarrow$ & FD $\downarrow$ & KL $\downarrow$ &IS $\uparrow$ &CLAP $\uparrow$ & OVL $\uparrow$ & REL $\uparrow$ \\

\midrule
\midrule
\multicolumn{10}{c}{\textit{Cascade (``W" refers to the WHISPER ASR model)}} \\
\midrule
\midrule

 W + AudioLDM 2 & 265.83 \spm{72.89}& 52.78 \spm{1.66} &2.40 &20.92 &1.62 & 8.38 & 0.377 & 3.54 \spm{1.02}  & 3.54 \spm{1.15}  \\

 W + TANGO 2 & 156.33 \spm{62.16} & 14.62 \spm{0.10} & 2.57 & 19.89 & \textbf{1.22} & 8.75 & \textbf{0.466} & 3.39 \spm{0.88} &3.78 \spm{1.12}  \\

 W + AUDIT & 187.86 \spm{75.75} & 9.13 \spm{0.43}& 6.20 & 41.89 & 1.88 & 6.26 & 0.293 &2.74 \spm{1.08}  &3.16 \spm{1.25}  \\

\midrule
\midrule
\multicolumn{10}{c}{\textit{End-to-End (Ours)}} \\
\midrule
\midrule

STAR   & 35.93 \spm{14.50} & \textbf{2.76} \spm{0.03} & \textbf{2.23} & \textbf{15.42} & 1.48 & \textbf{12.22} & 0.422 & \textbf{4.22} \spm{0.92} &\textbf{4.30} \spm{0.88}\\


\hspace{1em} w/o BridgeNet& \textbf{32.83} \spm{26.29} & 2.90 \spm{0.10} & 3.63 & 16.20 & 1.68 & 9.54 & 0.390 & - & -\\ 

\hspace{1em} DAC Encoder& 58.68 \spm{30.44} & 2.77  \spm{0.03} & 5.49 & 29.69 & 4.21 & 6.62 & 0.112 & - & -\\ 

\bottomrule
\end{tabular}
\label{tab:result}
\end{center}
\end{table*}

\textbf{Baseline} 
We incorporate cascade systems for comparison, combining ASR and TTA.
For ASR, we use WHISPER~\cite{radford2023robust}, while for TTA, we experiment several mainstream models: AudioLDM 2~\cite{liu2023audioldm2}, TANGO 2~\cite{majumder2024tango} and AUDIT~\cite{wang2024audit}. 
We adopt default settings for these models.
For STAR, we use an NFE of 20 and a guidance scale of $5.0$.

\textbf{Metrics} Common objective audio generation metrics, including FAD, FD, KL divergence, inception score (IS) and CLAP similarity, are employed.
We also perform human subjective evaluation, where evaluators listen to 10 samples from each model and score them on two dimensions: overall audio quality (OVL) and relevance to the input (REL).  
All evaluators are screened for no hearing loss and have university-level education from prestigious institutions.
They use designated headphones during evaluation.

\section{Results and Analysis}
We first evaluate the effectiveness of spoken sound event semantic representation extracted by different speech encoders and bridge networks.
Then, we compare STAR with cascaded baselines in terms of latency and generation performance, with ablations on the architecture.

\subsection{Representation Learning Evaluation}
\label{subsec:speech_rep_res}
The spoken sound event semantic representation capabilities are shown in \Cref{tab:hear}.
We also incorporate text encoders with the caption input as a topline since the input is the original caption.
Two text encoders are evaluated: audio-oriented CLAP~\cite{wu2023large}, and general-purpose Flan-T5~\cite{chung2024scaling}.
Both have been shown effective in TTA~\cite{liu2023audioldm2,majumder2024tango}.
Results show that semantic representations are comparable to the caption representation topline, indicating \textbf{the spoken sound event semantics can be effectively
and directly extracted from speech signals}.
Specifically, HuBERT performs slightly better than WavLM.
Both outperforms DAC, as they focus on semantic information, whereas DAC emphasizes acoustic details.
The Q-Former exhibits superior performance across different representations compared to MLP. 
Therefore, we adopt HuBERT and Q-Former in the following experiments by default.

\subsection{Speech Processing Latency}
The E2E system mitigates the latency introduced by the cascaded system. 
We measured the time required to extract semantic representation from each speech input. 
For the cascaded system, this includes the latency of the ASR model and text encoder, while for the E2E system, it includes the latency of the speech encoder and bridge network. 
\Cref{tab:result} presents the average results on test set.
The cascaded system requires an average delay of at least $156ms$, whereas STAR processes the speech signal directly with only $36ms$ latency ($\approx 76.9\%$ reduction), significantly reducing latency and holding the potential to enhance interactivity.

\subsection{Generation Performance}

Results of different STA systems are presented in \Cref{tab:result}. 
Both objective and subjective evaluation show that STAR outperforms all cascaded systems.
We observe that the ASR results contain inevitable errors while some TTA models are quite sensitive to the text input, resulting in unsatisfactory performance.
In contrast, as an E2E system, STAR takes speech embedding directly as the input to the generation model, mitigating the error accumulation problem.
Overall, the performance demonstrates the feasibility of E2E STA, to support more seamless human-machine interaction.

We further conducted ablation studies on the impacts of representation learning and speech encoder selection, as shown in the lower half of \Cref{tab:result}.
When we directly adopt HuBERT features as the input (w/o BridgeNet), the speech processing latency reduces slightly but the generation performance degrades noticeably.
This validates the effectiveness of the bridge network in mapping speech embedding to spoken sound event semantics.
Replacing HuBERT features with DAC results in much worse performance.
This corresponds to the linear probing results in \Cref{subsec:speech_rep_res}, indicating that STA prefers semantic features than acoustic features as only the speech content is the effective information for generation.

\section{Conclusion}
This work proposes STAR, an E2E STA system that leverages representation learning to create sound effects from spoken descriptions.
Through comprehensive representation learning experiments, we demonstrate that the combination of pre-trained speech encoders and a bridge network can extract sufficient semantic information for audio generation. 
Utilizing a Q-Former bridge network and a two-stage training strategy, STAR significantly reduces the processing latency while outperforming the cascaded baseline systems.
Ablation studies validate the necessity of our model architecture.
We hope STAR will contribute to the design of future E2E speech dialogue models, endowing them with omni generation abilities.

\bibliographystyle{IEEEtran}
\bibliography{refs}

\begin{thebibliography}{10}
\providecommand{\url}[1]{#1}
\csname url@samestyle\endcsname
\providecommand{\newblock}{\relax}
\providecommand{\bibinfo}[2]{#2}
\providecommand{\BIBentrySTDinterwordspacing}{\spaceskip=0pt\relax}
\providecommand{\BIBentryALTinterwordstretchfactor}{4}
\providecommand{\BIBentryALTinterwordspacing}{\spaceskip=\fontdimen2\font plus
\BIBentryALTinterwordstretchfactor\fontdimen3\font minus \fontdimen4\font\relax}
\providecommand{\BIBforeignlanguage}[2]{{%
\expandafter\ifx\csname l@#1\endcsname\relax
\typeout{** WARNING: IEEEtran.bst: No hyphenation pattern has been}%
\typeout{** loaded for the language `#1'. Using the pattern for}%
\typeout{** the default language instead.}%
\else
\language=\csname l@#1\endcsname
\fi
#2}}
\providecommand{\BIBdecl}{\relax}
\BIBdecl

\bibitem{zhang2023speechgpt}
D.~Zhang, S.~Li, X.~Zhang, J.~Zhan, P.~Wang, Y.~Zhou, and X.~Qiu, ``Speechgpt: Empowering large language models with intrinsic cross-modal conversational abilities,'' \emph{arXiv preprint arXiv:2305.11000}, 2023.

\bibitem{xu2025qwen2}
J.~Xu, Z.~Guo, J.~He, H.~Hu, T.~He, S.~Bai, K.~Chen, J.~Wang, Y.~Fan, K.~Dang \emph{et~al.}, ``Qwen2. 5-omni technical report,'' \emph{arXiv preprint arXiv:2503.20215}, 2025.

\bibitem{hsu2021hubert}
W.-N. Hsu, B.~Bolte, Y.-H.~H. Tsai, K.~Lakhotia, R.~Salakhutdinov, and A.~Mohamed, ``Hubert: Self-supervised speech representation learning by masked prediction of hidden units,'' \emph{IEEE/ACM TASLP.}, vol.~29, pp. 3451--3460, 2021.

\bibitem{chen2022wavlm}
S.~Chen, C.~Wang, Z.~Chen, Y.~Wu, S.~Liu, Z.~Chen, J.~Li, N.~Kanda, T.~Yoshioka, X.~Xiao \emph{et~al.}, ``Wavlm: Large-scale self-supervised pre-training for full stack speech processing,'' \emph{IEEE/ACM JSTSP.}, vol.~16, no.~6, pp. 1505--1518, 2022.

\bibitem{chu2023qwen}
Y.~Chu, J.~Xu, X.~Zhou, Q.~Yang, S.~Zhang, Z.~Yan, C.~Zhou, and J.~Zhou, ``Qwen-audio: Advancing universal audio understanding via unified large-scale audio-language models,'' \emph{arXiv preprint arXiv:2311.07919}, 2023.

\bibitem{kumar2024high}
R.~Kumar, P.~Seetharaman, A.~Luebs, I.~Kumar, and K.~Kumar, ``High-fidelity audio compression with improved rvqgan,'' \emph{Proc. NIPS}, vol.~36, 2024.

\bibitem{devlin2018bert}
J.~Devlin, M.-W. Chang, K.~Lee, and K.~Toutanova, ``Bert: Pre-training of deep bidirectional transformers for language understanding,'' in \emph{Proc. NAACL}, 2019, pp. 4171--4186.

\bibitem{liu2022flow}
X.~Liu, C.~Gong, and Q.~Liu, ``Flow straight and fast: Learning to generate and transfer data with rectified flow,'' \emph{arXiv preprint arXiv:2209.03003}, 2022.

\bibitem{chen2024f5}
Y.~Chen, Z.~Niu, Z.~Ma, K.~Deng, C.~Wang, J.~Zhao, K.~Yu, and X.~Chen, ``F5-tts: A fairytaler that fakes fluent and faithful speech with flow matching,'' \emph{arXiv preprint arXiv:2410.06885}, 2024.

\bibitem{hai2024ezaudio}
J.~Hai, Y.~Xu, Z.~H. Zhang, C.~Li, H.~Wang, M.~Elhilali, and D.~Yu, ``Ezaudio: Enhancing text-to-audio generation with efficient diffusion transformer,'' in \emph{Proc. Interspeech}, 2025, pp. 4233--4237.

\bibitem{kim2019audiocaps}
C.~D. Kim, B.~Kim, H.~Lee, and G.~Kim, ``Audiocaps: Generating captions for audios in the wild,'' in \emph{Proc. NAACL}, 2019, pp. 119--132.

\bibitem{kim2021conditional}
J.~Kim, J.~Kong, and J.~Son, ``Conditional variational autoencoder with adversarial learning for end-to-end text-to-speech,'' in \emph{International Conference on Machine Learning}.\hskip 1em plus 0.5em minus 0.4em\relax PMLR, 2021, pp. 5530--5540.

\bibitem{gemmeke2017audio}
J.~F. Gemmeke, D.~P. Ellis, D.~Freedman, A.~Jansen, W.~Lawrence, R.~C. Moore, M.~Plakal, and M.~Ritter, ``Audio set: An ontology and human-labeled dataset for audio events,'' in \emph{Proc. ICASSP}.\hskip 1em plus 0.5em minus 0.4em\relax IEEE, 2017, pp. 776--780.

\bibitem{radford2023robust}
A.~Radford, J.~W. Kim, T.~Xu, G.~Brockman, C.~McLeavey, and I.~Sutskever, ``Robust speech recognition via large-scale weak supervision,'' in \emph{Proc. ICML}.\hskip 1em plus 0.5em minus 0.4em\relax PMLR, 2023, pp. 28\,492--28\,518.

\bibitem{liu2023audioldm2}
H.~Liu, Q.~Tian, Y.~Yuan, X.~Liu, X.~Mei, Q.~Kong, Y.~Wang, W.~Wang, Y.~Wang, and M.~D. Plumbley, ``Audioldm 2: Learning holistic audio generation with self-supervised pretraining,'' \emph{arXiv preprint arXiv:2308.05734}, 2023.

\bibitem{majumder2024tango}
N.~Majumder, C.-Y. Hung, D.~Ghosal, W.-N. Hsu, R.~Mihalcea, and S.~Poria, ``Tango 2: Aligning diffusion-based text-to-audio generations through direct preference optimization,'' in \emph{Proc. ACM. MM}, 2024, pp. 564--572.

\bibitem{wang2024audit}
Y.~Wang, Z.~Ju, X.~Tan, L.~He, Z.~Wu, J.~Bian \emph{et~al.}, ``Audit: Audio editing by following instructions with latent diffusion models,'' \emph{Proc. NIPS}, vol.~36, 2024.

\bibitem{wu2023large}
Y.~Wu, K.~Chen, T.~Zhang, Y.~Hui, T.~Berg-Kirkpatrick, and S.~Dubnov, ``Large-scale contrastive language-audio pretraining with feature fusion and keyword-to-caption augmentation,'' in \emph{IEEE International Conference on Acoustics, Speech and Signal Processing}, 2023.

\bibitem{chung2024scaling}
H.~W. Chung, L.~Hou, S.~Longpre, B.~Zoph, Y.~Tay, W.~Fedus, Y.~Li, X.~Wang, M.~Dehghani, S.~Brahma \emph{et~al.}, ``Scaling instruction-finetuned language models,'' \emph{Journal of Machine Learning Research}, vol.~25, no.~70, pp. 1--53, 2024.

\end{thebibliography}

\end{document}